\documentclass[aps,pre,superscriptaddress,twocolumn,amsmath,amssymb,showpacs]{revtex4}
\usepackage{graphicx}
\usepackage{dcolumn}
\usepackage{bm}
\usepackage{color}
\begin{document}

\title{Disturbance accelerates the transition from 
low- to high- diversity state in a model ecosystem }

\author{Filippo Botta}
\affiliation{Center for Models of Life, 
Niels Bohr Institute, University of Copenhagen, Blegdamsvej 17, DK-2100 Copenhagen, Denmark}

\author{Namiko Mitarai}
\affiliation{Center for Models of Life, 
Niels Bohr Institute, University of Copenhagen, Blegdamsvej 17, DK-2100 Copenhagen, Denmark}

\date{\today}


\begin{abstract}
The effect of disturbance on a model ecosystem of 
sessile and mutually competitive species 
[Mathiesen {\it et al.} Phys. Rev. Lett. {\bf 107}, 188101 (2011);
Mitarai {\it et al.} Phys. Rev. E {\bf 86}, 011929 (2012)
]
is studied. The disturbance stochastically removes 
individuals from the system, and the created empty sites 
are re-colonized by neighbouring species. 
We show that the stable high-diversity state, 
maintained by occasional cyclic species interactions  
that create isolated patches 
of meta-populations,
is robust against small disturbance.
We further demonstrate that
finite disturbance can accelerate the transition from 
the low- to high-diversity state by helping 
creation of small patches through diffusion of 
boundaries between species with stand-off relation.
\end{abstract}



\maketitle


\section{Introduction}
The high-diversity of sessile species in 
a limited space,
observed in for example 
crustose lichen \cite{harris1996competitive,jettestuen2010competition}, 
coral reef \cite{buss1979competitive,volkov2007patterns} and epifaunal species \cite{kay1981occupation} ecosystems,
is one of the unsolved puzzle in ecology. 
Interestingly, these examples often 
lacks an obvious dominant species in their competition
\cite{harris1996competitive,jettestuen2010competition,buss1979competitive,kay1981occupation}. 
Instead,  many competitive equivalent species,
stand-offs (i.e. a boundary between two species stays still), and 
cyclic  (non-transitive) relationships are observed,
which should contribute the maintenance of the high biodiversity. 

Cyclic relations of species are extensively studied 
\cite{gilpin1975limit}, which gives coexistence
of oscillating populations for some time,  but in 
a well mixed system noise due to finite population 
eventually leads to the extinctions of the species and dominance of one or 
a few species \cite{Reichenbach2006},
unless the strength of the interaction between the species and the 
selection pressure is chosen in the range that allows coexistence  \cite{claussen2008cyclic}.
Recent work  \cite{knebel2013coexistence} provides the coexistence criteria of a given interaction network in well-mixed system with conservative dynamics.
By combining cycles with space to limit the interaction 
to local neighbors, the coexistence is found to be stabilized
\cite{boerlijst1991spiral, laird2006competitive,szabo2001defensive, czaran2002chemical,kerr2006local,perc2007noise,
lutz2012intransitivity, Roman2013interplay}.
For sessile species, 
the non-transitive relationships between species 
are found to 
prolong the the coexistence time than hierarchical relationships  \cite{buss1979competitive,
jackson1975alleopathy,karlson1984competition}.

Mathiesen {\sl et al.} \cite{Mathiesen-PRL2011} proposed  
 a spatial model ecosystem of sessile and mutually excluding organisms, 
 inspired by crustose lichen communities.
The model considers the overgrowth or allelopatic interaction between species,
that allows a species to invade a space occupied already by another species.   
Slow introduction of new species to the system is allowed, while stochastic extinction can also happen due to the finite population,  therefore the number of the 
species  in the system is not fixed {\it a priori} in the model \cite{culture}. 
In the limit of slow introduction rate, it was demonstrated that the model showed a discontinuous transition from the low- to high-diversity state 
as  the invasion interaction is reduced, hence the fraction of stand-offs increased. 
In the high-diversity state,  the spatial distribution of species is 
self-organized into fragmented patches, 
and species implicitly protect each other from the direct contact with the 
``competitively superior'' species
to allow stable coexistence. 
It has been demonstrated that cyclic relationships of length 4 to 6 are
necessary for maintenance of diversity 
through creation of patches 
\cite{Mitarai-2012}.

One of the biologically important extensions of the model 
is to include random disturbance from environment 
and/or natural deaths of the individuals. 
The disturbance creates empty space, which can
be recolonized from neighboring species \cite{karlson1984competition}.
Such a disturbance creates free space available for all species, 
and at the same time the species with the biggest population may be 
hit more often by disturbance. 
Therefore, disturbance may help inferior species to compete with dominant species
\cite{dayton1971competition,paine1974intertidal}.
When the disturbance is very  high, the species interaction becomes 
irrelevant for the ecosystem \cite{karlson1984competition}, 
and a large influx of new species 
becomes necessary to keep the diversity.
The biodiversity in a community obtained by balance between
the influx of new species and random  
extinction due to stochastic growth and death 
without species interaction  is discussed in the neutral theory  of biodiversity 
\cite{hubbell2001unified,volkov2003neutral,volkov2007patterns}.

Though the high-diversity state of the 
model in ref. \cite{Mathiesen-PRL2011}  is 
realized as the balance between the slow influx of new species
and the occasional extinction, it is 
different from the neutral theory situation because
the high-diversity state maintained in the small influx  limit 
relies on the spacial structure created by the species interactions.   
The effect of disturbance was tested before \cite{Mathiesen-PRL2011} 
by emptying a fraction of sites prior to the introduction of 
new species, and it was found that  high diversity is maintained as long 
as the removal is less that 10\%.
However,  the systematic study on either the stochastic disturbance 
on the dynamics or the relation between the 
disturbance and influx of new species has not been performed yet. 
Since the species interaction becomes irrelevant in  the large disturbance limit, 
our interest is in the effect of small but finite disturbance to the model behavior. 

In this paper, we study the ecosystem model  \cite{Mathiesen-PRL2011,Mitarai-2012}
with a finite disturbance rate. 
We show that, when the disturbance rate is small enough 
for a given introduction rate of new species, 
the clear distinction between the low and high-diversity states
and the sharp transition between them are maintained.
We further demonstrate that the disturbance can even enhance 
the transition from the low- to high-diversity state,
by mediating the fragmentation of the species into spatially separated patches. 

\section{Model}
\subsection{Model algorithm \label{ModelA}}
The ecosystem is modelled on a two-dimensional square lattice of the linear size $L$ with variable number of species,
where each site can either be empty or hold at most one species. 
The species interactions are characterized by a 
randomly assigned directed network,   
and the network connectivity is parametrized by the
probability  $\gamma$ (see below). 
The interaction takes place only 
among spatially neighboring species. 
New species are introduced with a constant rate $\alpha_N$ 
per site ($\alpha_N\times L^2$ per system),  while randomly chosen sites are emptied
 by disturbance with a rate $\omega_N$ per site ($\omega_N \times L^2$ per system).

Each update of our model consists of the following three possible events: 
\begin{enumerate}
\item {\it Introduction of new species.} 
With probability $\alpha_N$, 
choose a random lattice site $j$.
(a) If the site $j$ is empty, 
then a new species $s$ is introduced at the site $j$,
and assigned random interactions $\Gamma (s,u)$ and 
 $\Gamma(u,s)$ with all existing species $u$ in the system.
Each of these interactions are assigned value 1 with probability $\gamma$, 
and otherwise set to $0$.
$\Gamma(s,u)=1$ indicates that the species $s$ can invade the species 
$u$, while $s$ cannot invade $u$ if $\Gamma(s,u)=0$  (See invasion rule below).
(b) If instead the site $j$ is occupied already by another species $v$, then a new species $s$ is introduced with  probability $\gamma$, which is the probability that 
the species $s$ can invade the species $v$.
When $s$ is introduced,  $\Gamma (s,u)$ and  $\Gamma(u,s)$ 
for all existing species $u$ in the system
are assigned in the same way 
as (a), and then $\Gamma(s,v)$ is set to $1$.
\item {\it Disturbance.} With probability $\omega_N$, choose a 
lattice site $i$
randomly. If there exist a species on the site $i$, 
make the site $i$ empty, hence the population of the species will 
be reduced by one.
\item 
{\it Invasion.}
Choose a random lattice site $i$. 
If there is a species $s(i)$ at the site $i$, 
choose one of its 4 neighbor sites $j$ randomly. 
If the site $j$ is empty, 
or if there is a species $s(j)$ at the site $j$
and $s(i)$ can invade $s(j)$ (i.e. $\Gamma[s(i),s(j)]=1$),
then the site $j$ will be updated to be occupied by
the species $s(i)$ by setting $s(j)=s(i)$.
\end{enumerate}

One time unit is defined as $L^2$ repeats of the procedures 1 to 3.
Therefore, per time unit, on average each site makes one attempt to invade a neighbor, 
$\alpha_N \times L^2$ new species attempt to enter the system,
and $\omega_N \times L^2$ individuals are removed. 
Since $\alpha_N$ and $\omega_N$ are defined as a probability 
for the introduction and the death occur per time unit, respectively, 
they can take any value between 0 and 1. 
$\omega_N=0$ recovers the original ecosystem model 
in \cite{Mathiesen-PRL2011,Mitarai-2012}.

Because of the random assignment of the 
interaction matrix $\Gamma$, 
there will be no dominant species.
For a given pair of species $(s,u)$, 
one of them can be dominant ($\Gamma(s,u)=1$
and $\Gamma(u,s)=0$ or {\it vice versa}),
or they can be competitively equivalent.
Note that there are two kinds of 
competitive equivalence: it can be 
with active invasion to each other 
($\Gamma(s,u)=\Gamma(u,s)=1$), or
with the stand-off relation ($\Gamma(s,u)=\Gamma(u,s)=0$).
The stand-off relation increases for smaller $\gamma$, 
mediating the stable coexistence of many species when $\omega_N=0$. 

\subsection{Simulation setup}
In the following simulations, we use $L=200$ under periodic boundary condition.
System size dependence was was studied before
at $\omega_N=0$ \cite{Mathiesen-PRL2011} and will be  
summarized in subsection \ref{summary0}.
$L=200$ was found to be enough to observe 
the transition to the high-diversity state. 
Initial condition is an empty system, unless otherwise noted. 

In order to reduce the computation time, 
some of the simulations with small values of $\alpha_N$ and $\omega_N$ 
were performed by the event-driven type algorithm, 
where the possible events are listed and time to the next event was 
drawn accordingly from the exponential distribution.
This gives statistically the same results as the described 
random sequential updates.

\section{Results}
\begin{figure}[t]
\includegraphics[width=0.45\textwidth]{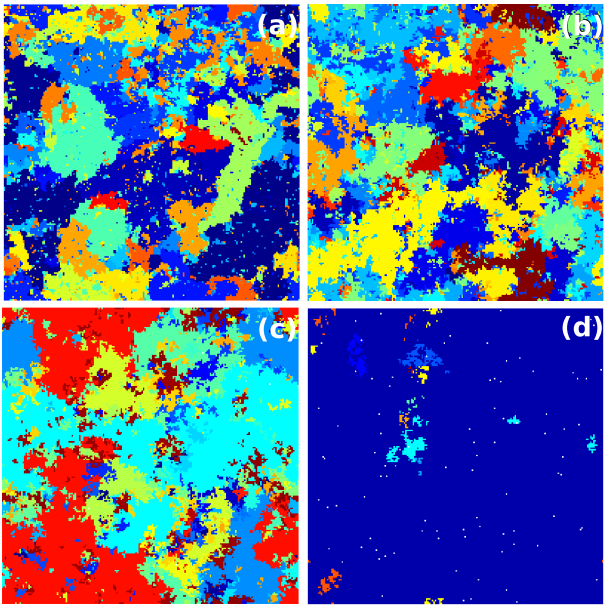}
\caption{ (color online)
Snapshots of in the high-diversity state with 
$\alpha_N=2.5\times 10^{-7}$,
$\gamma=0.03$, with various death rate $\omega_N$.
(a) $\omega_N=0$,
(b) $\omega_N=2.5\times 10^{-5}$,
(c) $\omega_N=2.5\times 10^{-4}$,
(d) $\omega_N=2.5\times 10^{-3}$.
The empty sites are shown as white, and the sites with species are 
filled.
}
\label{Snapshots}
\end{figure}

\subsection{Summary of the behavior without the disturbance
\label{summary0}}
We first summarize the relevant findings 
in the previous papers
for no disturbance ($\omega_N=0$)  
\cite{Mathiesen-PRL2011,Mitarai-2012}, to clearly demonstrate the effect of 
the disturbance later.

In the no-disturbance case, the time-averaged diversity (number of species in the system) 
$\langle D\rangle$ was found to decrease to one as $\alpha_N\to 0$ 
for $\gamma>\gamma_c\approx 0.06$, 
while  for $\gamma<\gamma_c$,
$\langle D\rangle$ converges to a finite non-zero value as 
$\alpha_N\to 0$. 
In the limit of $\alpha_N \to 0$, or the quasistatic version of the model where a new species is inserted into the system only after all the activity in the system has stopped \cite{quasistatic},
the state with the diversity $D=1$ is an absorbing state, since newly 
added species will simply replace the previous species.  
When the simulation is started from the high diversity state in the quasi-static limit, the system stably remains in the high-diversity state for $\gamma<\gamma_c$, while for $\gamma>\gamma_c$ the diversity goes down to one. Namely,  the diversity shows 
a discontinuous transition as a function of $\gamma$ in the 
$\alpha_N\to 0$ limit. 

When $\alpha_N$  is finite but small enough ($\alpha_N\le 2.5\times 10^{-7}$), 
the high-diversity state is mono-stable for 
$\gamma<\gamma_c$, while the diversity $D$ shows clear bi-stability between the high-diversity state and the low-diversity state for $\gamma>\gamma_c$, and hence the transition in the long-time average $\langle D\rangle$ appears softer as a function of $\gamma$ for non-zero $\alpha_N$. 
When $\alpha_N$ is large,  the difference between the high- and the low-diversity state is smeared out even without taking long-time average, since the high introduction of species forces the system to contain many species all the time.  
 
 A snapshot of the system in the high-diversity state 
 with $\omega_N=0$, $\alpha_N=2.5\times 10^{-7}$, $\gamma=0.03$ for $L=200$ is 
 shown in Fig.~\ref{Snapshots}a.
 We can see that the species are separated into many patches
created by species interactions. 
When $\gamma<\gamma_c$, the system stabilizes to the situation where 
most of the species cannot invade their neighbors, and slow introduction 
of new species gives local updates of the species distribution. 
 The spatial separation due to the patch formation implicitly protect each other from the ``competitively superior'' species
 to allow the high diversity in the system. 
 
The maximum patch-size has a well-defined cut-off around $10^4$ sites in area for small enough $\alpha_N$ for $\gamma<\gamma_c$, for $L\ge 200$. 
If the total system size $L$ is smaller 
than $\approx 150$, the high-diversity state is not stable 
because the large patches interfere with the system size, 
and we do not observe the transition. For a large enough system size, 
the transition point $\gamma_c$ does not depend on $L$, and 
the diversity increases linearly with the total system's area, $L^2$. 

We have also shown \cite{Mitarai-2012}
that cyclic relationships, especially of the 
length 4 to 6, are needed to maintain the high diversity.
Here, for example cyclic relation of length 4 means
the relation $A \to B \to C\to D \to A$,
where $X \to Y$ represents that the species $X$
can invade the species $Y$.
Such cyclic relationships give complex spatiotemporal dynamics,
and when some of the species go extinct due to the stochasticity,
many patches can be left behind.
In the case of cycle of length 4,  for example, 
if A dies out first, it is likely that B displaces C
before C can displace D since $B$ will not be attacked any more; 
in the end patches of B and D will be left and they will coexist
with stable boundary between them.
Obviously the cycles of length 2 and 3 do not leave patches, 
while longer cycles are less likely to be activated.
It has been shown that, when $\omega_N=0$,
the high-diversity state is not stable
without cycles of length $4-6$ in the $\alpha_N\to 0$ limit
for $\gamma=0.025$ \cite{Mitarai-2012},
suggesting the necessity of the patch creation through the cyclic relations.

 \subsection{Effect of the the disturbance}
 \subsubsection{Average diversity}
 \begin{figure}[h]
\includegraphics[width=0.45\textwidth]{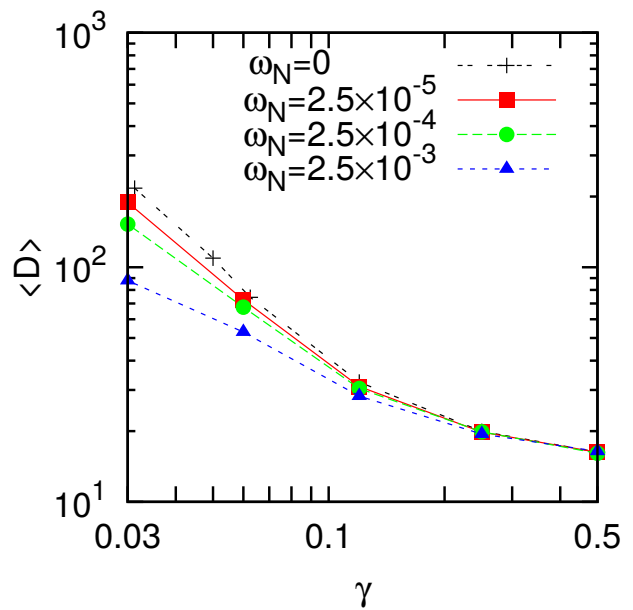}
\caption{(color online)  Long-time average of the diversity $\langle D\rangle$
as a function of $\gamma$.
$\alpha_N=2.5\times 10^{-6}$, and results for $\omega_N=0, 2.5\times 10^{-5}, 
2.5\times 10^{-4}, $ and $2.5\times 10^{-3}$
are shown. 
The average was taken from the data of the duration $9\times 10^6$ time steps or longer.
The standard error is smaller than the symbol size.
}\label{a0c1}
\end{figure}
\begin{figure}[h]
\includegraphics[width=0.45\textwidth]{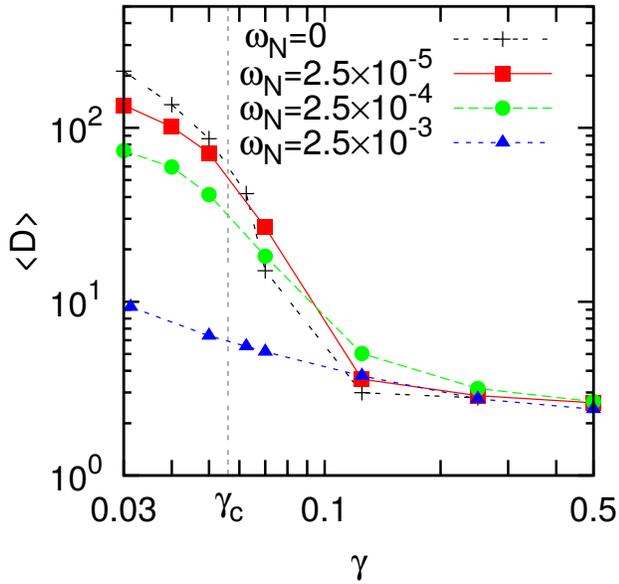}
\caption{(color online) Long-time average of the diversity $\langle D\rangle$
as a function of $\gamma$. 
$\alpha_N=2.5\times 10^{-7}$, and results for $\omega_N=0, 2.5\times 10^{-5}, 
2.5\times 10^{-4}, $ and $2.5\times 10^{-3}$
are shown. 
The average was taken from the data longer than the following duration;
For $\omega_N=0$, $\gamma\le 0.125$:  $7\times 10^8$ time steps.
For $\omega_N=0$, $\gamma= 0.25$:  $1\times 10^8$ time steps.
For $\omega_N=0$, $\gamma= 0.5$:  $1.9\times 10^7$ time steps.
For $\omega_N=2.5\times 10^{-5}$, $\gamma\le 0.25$:  $1\times 10^8$ time steps.
For $\omega_N=2.5\times 10^{-5}$, $\gamma= 0.5$:  $8\times 10^7$ time steps.
For $\omega_N=2.5\times 10^{-4}$, $\gamma\le 0.125$:  $1\times 10^8$ time steps.
For $\omega_N=2.5\times 10^{-4}$, $\gamma= 0.25$:  $9\times 10^7$ time steps.
For $\omega_N=2.5\times 10^{-4}$, $\gamma= 0.25$:  $5\times 10^7$ time steps.
For $\omega_N=2.5\times 10^{-3}$, $\gamma\le 0.075$:  $9\times 10^7$ time steps.
For $\omega_N=2.5\times 10^{-3}$, $\gamma= 0.25, 0.5$:  $9\times 10^6$ time steps.
The standard error is smaller than the symbol size.
}
\label{averageD}
\end{figure}
Now we move on to the case with the  non-zero disturbance rate, $\omega_N$.
 Figures~\ref{Snapshots}b-d show snapshots  for 
 $\alpha_N=2.5\times 10^{-7}$ and $\gamma=0.03$ 
 with $\omega_N=2.5\times 10^{-5}, 2.5 \times 10^{-4}$, and $2.5\times 10^{-3}$, 
 respectively, in the high-diversity state. 
When the disturbance happens at the boundary between two stand-off species, 
 the empty sites will be filled by one of the neighboring species, resulting in slow 
 diffusion of the boundary. 
 This situation is the same as the  voter model \cite{dornic2001critical},
 where a smooth interface roughens due to the lack of the surface tension. 
 This makes the boundary between species more 
 rugged and leads to elimination of the patches as $\omega_N$ increases. 
  
We then studied the $\alpha_N$, $\gamma$, and $\omega_N$ dependence of 
the time averaged diversity $\langle D\rangle$. 
With finite $\omega_N$ and large enough $\alpha_N \ge 1.25 \times 10^{-6}$,
$\langle D\rangle$ decrease with increasing $\omega$ from zero
for any fixed value of $\gamma$.
An example is given in Fig.~\ref{a0c1} for $\alpha_N=2.5\times 10^{-6}$ 
with various values of $\omega_N$. This $\alpha_N$ is large enough to 
dominate the system's diversity for $\omega_N=0$ case
to smear out the difference between the high-diversity state 
and the low-diversity state. 
The monotonic dependence on $\omega_N$ suggest that  
$\omega_N$ is simply counteracting the diversity 
by accelerating the extinction through random death of individuals.

When $\alpha_N=2.5\times 10^{-7}$ and 
$\gamma< \gamma_c$, $\langle D\rangle$ again decreases monotonically with 
increasing $\omega_N$ as shown Fig.~\ref{averageD}.
This is expected because the disturbance results in the 
fluctuation of the stable boundary between species with stand-off relation, 
as depicted in Fig.~\ref{Snapshots}, namely 
too large $\omega_N$ destabilize the high-diversity 
state.  When $\gamma=0.03$, for example, the system
is monostable in the high-diversity state for $\omega_N < 2.5\times 10^{-4}$ 
over $10^8$ time steps, 
but it shows bistability between the high- and low-diversity states
when $\omega_N$ exceeds $\omega_N < 5\times 10^{-4}$.
Destabilization of the high-diversity state for $\gamma<\gamma_c$
also occurs when $\omega_N$ is kept finite
and $\alpha_N$ is decreased. 
For $\gamma=0.03$ and $\omega_N=2.5\times 10^{-5}$, 
the high-diversity state was stable for $\alpha_N=2.5\times 10^{-7}$
over $10^8$ time steps, but 
became clearly unstable when $\alpha_N=2.5\times 10^{-9}$.

\begin{figure}[t]
\includegraphics[width=0.45\textwidth]{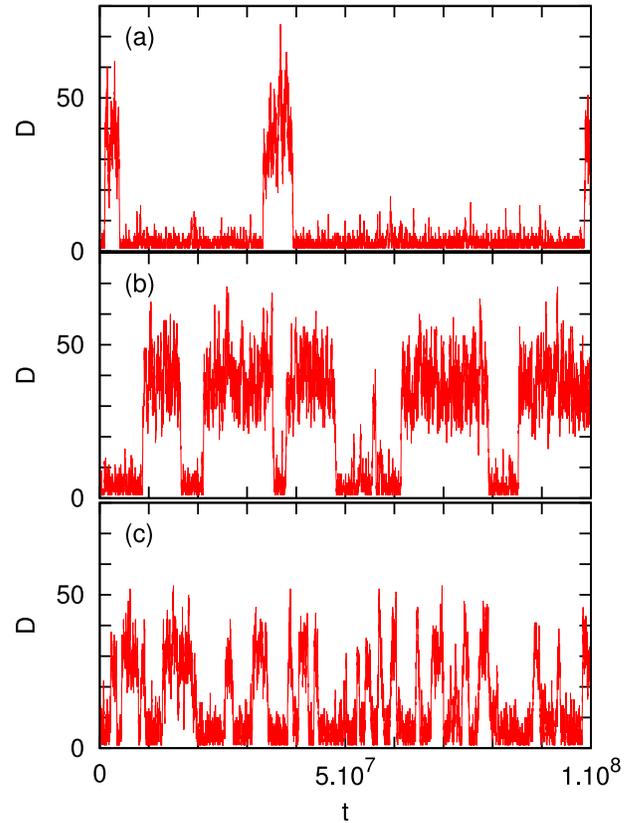}
\caption{(color online) Time dependence of diversity $D$ 
in the bistable region, $\alpha_N=2.5\times 10^{-7}$ and $\gamma=0.07$.
(a)$\omega_N=0$, (b)$\omega_N=2.5\times 10^{-5}$,(c)$\omega_N=2.5\times 10^{-4}$. 
The frequency to reach the high-diversity state increases with $\omega_N$,
while the value of the diversity in the high-diversity state
weakly decreases with $\omega_N$. 
}\label{timedep}
\end{figure}

Interestingly, we found that $\langle D\rangle$ shows 
a non-monotonic dependence on $\omega_N$ for 
$\alpha_N=2.5\times 10^{-7}$ and $\gamma>\gamma_c$ (Fig.~\ref{averageD}).
Especially, when $\gamma$ is just above $\gamma_c$, where the 
$\omega_N=0$ systems show bistability between the high- 
and low-diversity states, 
$\langle D\rangle$ increases when $\omega_N$ 
is increased from zero to $2.5\times 10^{-5}$ and then decreases with 
increasing $\omega_N$ further.

\subsubsection{Transition rates between the low-diversity state and 
the high-diversity state}
The non-monotonic dependence of the time-averaged diversity 
$\langle D\rangle$ on $\omega_N$ is 
due to the difference in the transition rates between the low- and 
high-diversity states.  As shown in Fig.~\ref{timedep} for 
$\alpha_N=2.5\times 10^{-7}$ and $\gamma=0.07$,
increasing $\omega_N$ slightly decrease the value of $D$ in the high-diversity state, but the effect is very weak. 
However, the transition rates for both from the low- to high-diversity state 
and the high- low-diversity state 
are significantly increased with increasing $\omega_N$. 
Especially, 
the enhancement of the transitions from the low- to high-diversity state is 
significant from $\omega_N=0$ case to the $\omega_N=2.5\times 10^{-5}$ case, 
which gives larger probability to be in 
the high-diversity state resulting in the higher value for $\langle D \rangle$.

\begin{figure}[t]
\includegraphics[width=0.45\textwidth]{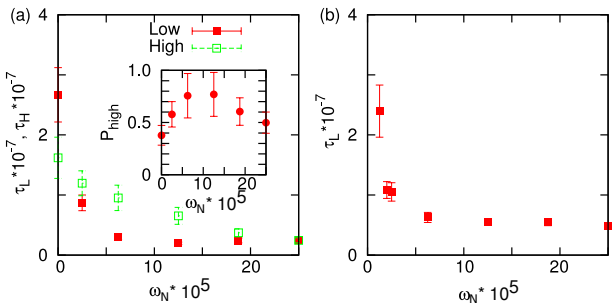}
\caption{(color online)
(a) Average transition time from the low- to high-diversity state $\tau_{L}$
and from the high- to low-diversity state $\tau_{H}$ 
as a function of $\omega_N$ 
in the bistable region, $\alpha_N=2.5\times 10^{-7}$, $\gamma=0.07$.
The transition between the high- and low-diversity states by using two threshold for the diversity: $D_l = 1$ as threshold for low diversity and $D_h(\omega_N)$ as the threshold for high diversity.
In order to determine $D_h(\omega_N)$,  we first define the high-diversity state as 
the state where $D$  has distinctly high diversity $D>1$ for longer than $10^5$ time steps,
and then calculated $D_h(\omega_N)$ as the average over the diversity over this period.
We then define a switching event from the low- to high-diversity state when $D$ exceeds $D_h$, while the reverse switching happens when $D$ reaches $D_l$.  
The life time of the low- (high-) diversity state
$\tau_{L}$  ($\tau_{H}$)
 is defined as the time between the high to low (low to high) switching event
and the low to high (high to low) switching  event \cite{comment}.
Inset: Probability to be in the high-diversity state
$P_{high}\equiv \tau_{H}/( \tau_{H}+\tau_{L})$, 
which has maximum at $\omega_N \approx 10^{-4}$.
(b) Average transition time from the low- to high-diversity state $\tau_{L}$
as a function of $\omega_N$ with $\alpha_N=2.5\times 10^{-7}$ and $\gamma=0.03$,
where the high-diversity state is monostable within the simulated timescale.
In this case, we start from $D=0$ and measure the time 
it takes before $D$ reaches the high diversity state value $D_h$. 
}
\label{transition}
\end{figure}
\begin{figure}[h]
\includegraphics[width=0.45\textwidth]{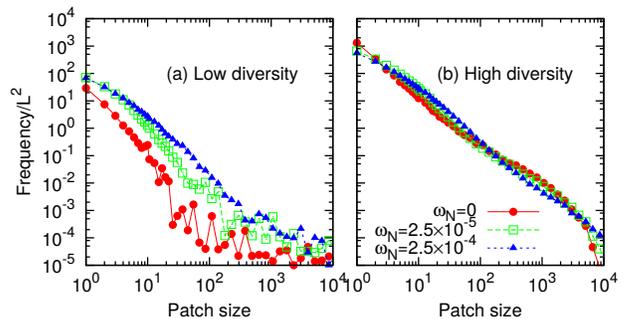}
\caption{(color online)
Patch size distribution for $\alpha_N=2.5\times 10^{-7}$ and $\gamma=0.07$
with $\omega_N=0, 2.5\times 10^{-5}, $ and $2.5\times 10^{-4}$.
(a) For low-diversity state.
(b) For high-diversity state.
}\label{patchdist}
\end{figure}

We quantify the average life time of the low- (high-) diversity state $\tau_{L}$  
($\tau_{H}$)  in the bistable regime as a function of $\omega_N$ (Fig.~\ref{transition}a). 
We see that $\tau_{L}$ quickly decreases about 9 fold when $\omega_N$ 
increases from zero to $5\times 10^{-5}$, and then saturate.
$\tau_{H}$ also decrease with $\omega_N$, but the change is slower. 
This results in a peak of the probability to be in the 
high-diversity state $P_{high}=\tau_{H}/(\tau_{H} +\tau_{L} )$,
leading to the non-monotonic dependence of $\langle D\rangle$ to $\omega_N$. 

The faster transition from the low- to high-diversity state 
is also seen in the monostable high-diversity regime. 
Figure~\ref{transition}b shows $\tau_{L}$ with $\alpha_N=2.5\times 10^{-7}$,
 $\gamma=0.03$,  for $0<\omega_N\le 2.5\times 10^{-4}$. Within this parameter range, 
 the high-diversity state is monostable over at least $10^8$ time steps.
 We measured the life time to the low-diversity state $\tau_L$ by
 starting from an empty system and averaged over at least 20 events.
 The resulting life time of the low-diversity state $\tau_{L}$
 shows again a sharp drop as $\omega_N$ increases from 0 to $\sim 5\times 10^{-5}$ and 
 then converges to a constant number as $\omega_N$ further increases. 
 
 We hypothesize that this acceleration of the transition to the high 
diversity by $\omega_N$
 is because disturbance-induced fluctuation of the boundaries
 creates more patches or meta-populations. 
 Suppose that the diversity is low, but still by chance a few species
 are coexisting,  each of them occupying one patch (a connected region)
 in the  system and they cannot invade each other. The boundaries between species 
 will diffuse due to the disturbance,   creating more and more small isolated patches.
 These additional patches allow the 
 system to host more species as species in a patch is replaced with a new species, 
 helping the system to reach high-diversity state.

 Figure~\ref{patchdist}a shows the patch-size distribution 
 for $\alpha_N=2.5\times 10^{-7}$ and $\gamma=0.03$, obtained from the 
 snapshots before the system reaches the high-diversity state. 
 We clearly see that the systems with non-zero $\omega_N$ have 
 more patches than the system with $\omega_N=0$, especially many of small patches with size $1$ to $10$.  In the high-diversity state for the system with 
the same parameters  (Fig.~\ref{patchdist}b), on the other hand, we see that 
 more and more patches of size 1 are removed as $\omega_N$ increases,
which contribute to the destabilization of the high-diversity state 
by $\omega_N$.
 
 \subsection{Necessity of the cyclic relationship for the high-diversity state} 
As summarized in subsection \ref{summary0},
it has been shown \cite{Mitarai-2012}
that, for $\omega_N=0$, cyclic relationships, especially of the 
length 4 to 6, are needed to maintain the high diversity.
We here investigate whether the patch creation by disturbance is enough 
or the system needs further creation of the patches through species 
interactions to keep the system in the high-diversity state. 

\begin{figure}[t]
\includegraphics[width=0.45\textwidth]{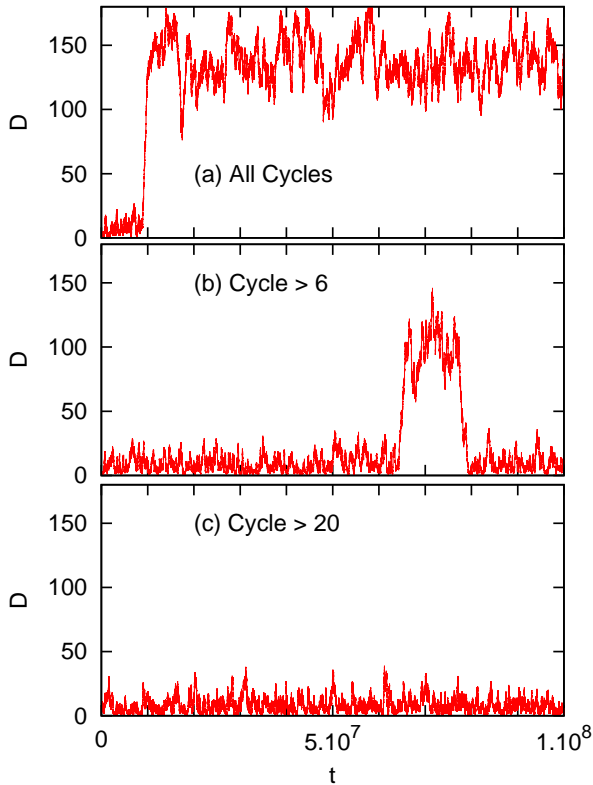}
\caption{ (color online)
Effect of cycles for $\alpha_N=2.5\times 10^{-7}$ and $\gamma=0.03$
with $\omega_N=2.5\times 10^{-5}$ on the time dependence of $D$.
(a) All cycles lengths are allowed ($C_{mim}=0$).
(b) Cycles of length longer than 7 are allowed ($C_{mim}=6$).
(c) Cycles of length longer than 20 are allowed ($C_{mim}=20$).
}\label{cycle}
\end{figure}

In order to test this,
we performed simulations with cycles of various degree by the following way \cite{Mitarai-2012}: 
When a new species was introduced, the corresponding entries in the interaction matrix 
$\Gamma$ were determined according to the given value of $\gamma$ , but if it would result in a cyclic relationship of length less than $C_{min}$, the species was rejected, and another new species was introduced, which again was assigned random interactions according to $\gamma$. 

Figures \ref{cycle}(a), (b), and (c) show 
the time series of the diversity $D$ for 
$C_{min}=0$ (i.e., the original model with all the cycles),
$C_{min}=6$, and $C_{min}=20$, respectively, 
with $\alpha_N=2.5\times 10^{-7}$, $\gamma=0.03$, and $\omega_N=2.5\times 10^{-5}$. 
When all the cycles are allowed, the transition 
to the high $D$  state occurs rather fast and it is 
monostable (Fig.\ref{cycle}a). 
The system can still go to the high-diversity 
state when only the cycles of length longer than 7 are allowed,
but the state is not stable (Fig.\ref{cycle}b).
When only the cycles of length longer than 20 are allowed, 
the transition to high-diversity state did not happen within 
$2\times 10^8$ 
time steps (Fig.\ref{cycle}c shows only half of the time series).

\section{Summary}
\begin{figure}[t]
\includegraphics[width=0.45\textwidth]{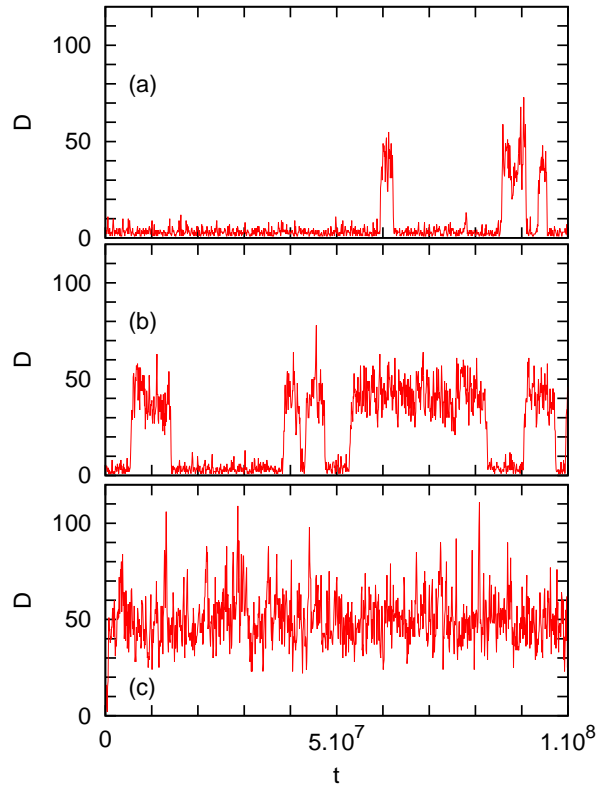}
\caption{(color online) Time dependence of diversity $D$ 
in the bistable region, $\alpha_N=2.5\times 10^{-7}$ and $\gamma=0.07$,
with various mobility rate $\mu_N$. The mobility is defined by replacing
the disturbance step 2. in the algorithm in section \ref{ModelA} 
with the following step:
{\it 2'. Mobility. With the probability $\mu_N$, 
select a neighboring pair of sites randomly, and exchange the species 
between the chosen sites.}
(a)$\mu_N=2.5\times 10^{-8}$, (b)$\mu_N=2.5\times 10^{-7}$, (c)$\mu_N=2.5\times 10^{-6}$. 
The frequency to reach the high-diversity state increases with $\mu_N$,
while the transition to the low diversity state is suppressed
with increasing $\mu_N$. 
}\label{mobility}
\end{figure}

We studied a model ecosystem with a finite rate of disturbance.
The transition to the high-diversity state 
in the ecosystem model \cite{Mathiesen-PRL2011,Mitarai-2012}
is shown to be robust against the finite disturbance rate $\omega_N$;
as long as $\omega_N$ is small enough compared to 
the introduction rate of the new species $\alpha_N$,
the clear separation between the high-diversity state 
and the low-diversity state is observed, and 
the high-diversity state is stable over long time for small enough $\gamma$.

We demonstrated that, even though the disturbance 
lowers the diversity at the high-diversity state, 
it can accelerate the transition from the low- to 
high-diversity state.  We also found that 
the cyclic interactions of proper lengths are still necessary 
for both the stable high-diversity state 
and the acceleration of transition to the high diversity by $\omega_N$.
These results may understood as follows:
The disturbance results 
in the diffusion of the boundary between 
species with stand-off relation, 
which creates small patches that can host more species.
The patches created by the disturbance at the low-diversity state accelerate 
the transition to the high-diversity state
by increasing the chance of cycles 
activated in the system.
In this way small disturbance can promote the 
high-diversity state.

 The effect of the disturbance studied here 
may be expected to be close to the effect of local mobility 
\cite{reichenbach2007mobility}
of two neighboring individuals able to swap their position with a small 
probably, because both makes a boundary between species diffuse. 
Preliminary result of the model with small $\alpha_N$ with local mobility 
showed the enhancement of the transition to the high-diversity state for the 
small mobility (Fig.~\ref{mobility}).  
However, there is a qualitative differences between the mobility and 
the disturbance: the mobility does not directly decrease the population of
each species. Therefore, the small mobility does 
not enhance the transition from the high-diversity state 
to the low-diversity state.
It is also known that the mobility will enhance the formation of 
"defensive alliances", by promoting the segregation of the species 
so that the ones without direct competition appears more often next to each other \cite{szabo2001defensive, Roman2013interplay}. 
This is also expected to promote the high diversity by reducing the 
extinction due to species interactions. 
At the same time, we have demonstrated in the 
previous paper \cite{Mathiesen-PRL2011} that 
the random neighbor version of the model, 
where the invasion step happens between 
two randomly chosen sites irrespective of the distance between them, 
does not support the high diversity state.  
The random neighbor situation should correspond to the high mobility limit, 
therefore non-monotonic dependence of the diversity 
on the mobility may be observed. This should be clarified in the future work.

The mobility in the longer distance, on the other hand,  is also
natural to take into account, when considering
 lichen spreading spores via wind   \cite{munoz2004wind}
or coral spreading with the water circulation \cite{wolanski1989trapping}.
In the present model, the parameter $\alpha_N$ is interpreted as immigration of 
a new species from an external environment. It will be interesting to compare 
the present model and the model with global mobility in a large system size. 

\acknowledgments
NM thanks Kim Sneppen for fruitful discussions.
This work has been supported by The Danish National Research Council,
through Center for Models of Life.


\bibliography{ecosystem}

\begin{thebibliography}{34}
\bibitem [{com()}]{comment}%
  \BibitemOpen
  \href@noop {} {}\bibinfo {note} {Because of the stochasticity of the
  dynamics, it is necessary to define the two thresholds to distinguish the
  high- and the low-diversity states. We checked that the results are not
  sensitive to exact definition of the threshold by increasing $D_l$ to 3 and
  decreasing $D_h(\omega)$ to 90\% of the original value.}\BibitemShut {Stop}%
\end{thebibliography}

\end{document}